\documentclass[11 pt]{article}
\usepackage{amsmath}
\usepackage{amsthm}
\usepackage{amssymb}
\usepackage{graphicx}
\usepackage{hyperref}
\usepackage{color}
\usepackage{booktabs}

\numberwithin{equation}{section}

\begin{document}

\title{Analysis of stock index with a generalized BN-S model: an approach based on machine learning and fuzzy parameters}
\author{Xianfei Hui\footnote{School of Management, Harbin Institute of Technology, Harbin, 150001, China; and Department of Mathematics, North Dakota State University,
 Fargo, North Dakota 58108, USA}, Baiqing Sun\footnote{School of Management, Harbin Institute of Technology, Harbin, 150001, China}, Hui Jiang\footnote{College of Management and Economics, Tianjin University, Tianjin 300072, China}, Indranil SenGupta \footnote{Corresponding author: indranil.sengupta@ndsu.edu}\footnote{Department of Mathematics, North Dakota State University, Fargo, North Dakota 58108, USA} }
\date{\today}
\maketitle

\begin{abstract}
In this paper we implement a combination of data-science and fuzzy theory to improve the classical Barndorff-Nielsen and Shephard model, and implement this to analyze the S\&P 500 index.
We pre-process the index data based on fuzzy theory. After that, S\&P 500 stock index for the past ten years are analyzed, and a deterministic parameter is extracted using various machine and deep learning methods. The results show that the new model, where fuzzy parameters are incorporated, can incorporate the long-term dependence in the classical Barndorff-Nielsen and Shephard model. The modification is based on only a few changes compared to the classical model. At the same time, the resulting analysis effectively captures the stochastic dynamics  of stock index time series. 
\end{abstract}



\section{Introduction}

Stock index, the most important indicator that reflects and predicts financial market fluctuations and global economic changes, is playing an increasingly important role in capital market.
The total value and trading volume of global trading products of stock index are increasing year by year. It is often used as a benchmark to measure the performance of investment funds, the basis of passive management that replicates its performance, derivative instruments involved in transactions, benchmark indicators for financial contracts, and refined risk management tools (see \cite{Laurent}). Most collective investment funds are based on various indices (see \cite{Petry}).

Many problems in the research of stock indices are closely related to the volatility of option prices (see \cite{Ruanx}). An effective tool to explore the law of asset volatility is stochastic volatility, which is the core variable of asset pricing, investment portfolio, and risk management. In the recent literature, the Barndorff-Nielsen and Shephard (BN-S) model is an effective and operable stochastic model, which is widely used in the analysis of financial asset prices in derivatives and commodity markets (see \cite{Barndorff,  barndorffn, Issaka}). With tractable mathematical properties, the BN-S model assumes that the return variance process obeys the non-Gaussian Ornstein-Uhlenbeck (OU) process. This can accurately describe the stochastic volatility of assets. 
The classical BN-S model has many advantages, such as, in some cases it is very effective and simple to use. However, it has many other problems, such as the lack of long-term dependence (see \cite{sengupta4}).

In recent literature, the properties and applications of the classical BN-S model and its extended models have been well studied. Benth et al. (see \cite{Benth}) hold that the forward price in the commodity market will jump according to the changes in the volatility process. Consequently, the non-Gaussian stochastic volatility model of the BN-S type, can be implemented to derive the spot price. SenGupta et al. (see \cite{Sengupta1}) improved the BN-S model from the perspective of the long-term dependence of derivative asset time series, and effectively used the new model to price European options and fitted the implicit volatility smile. Kallsen et al. (see \cite{Kallsenj}) considered jumps, stochastic fluctuations, and leverage effects in the application of the BN-S model, and derived option pricing according to the quadratic change of the stock price process. Jawadi et al. (see \cite{Jawadif}) implemented the BN-S model to decompose the intra-day volatility into continuous volatility and jumping volatility, and studied the relationship between the trading volume and volatility of different international stock markets. Habtemicael et. al. (see \cite{Habtemicaels}) proposed an optimized BN-S model of superimposed L\'evy processes driven by $\Gamma$ and inverse Gaussian distributions. Considering the possibility of two-way jumps in the asset price process, Bann{\"o}r et al. (see \cite{Bannor}) proposed an extended form of the BN-S stochastic volatility model. This is known as the double-sided BN-S model. Based on the stationary and self-decomposable distribution of the variance process, Awasthi et al. (see \cite{Awasthis}) provided an approximate expression of the BN-S model, and analyzed the first exit time and its distribution for a superposition of Brownian motion and L\'evy subordinator. Salmon et al. \cite{Salmon} improved the BN-S model based on the fractional Brownian motion. In another recent paper, Lin et al. (see \cite{Lin}) effectively implemented a BN-S type model for portfolio optimization. 

On the other hand, data-science, especially the machine/deep learning is affecting the financial market analysis in a a very productive way (see \cite{culkinr}). In innovative research on topics such as price prediction, risk control, volatility simulation, quantitative trading, data processing, and trend analysis, machine/deep learning technology is effectively implemented in recent years. For example, by applying deep learning methods to high-frequency financial data sets, Sirignano et al. (see \cite{sirignanoj}) found non-parametric evidence for the existence of general and fixed price formation mechanisms related to price dynamics. Cont et al. (see \cite{contr}) proposed a machine learning random algorithm, calculated the best trading strategy, and studied the sensitivity of the solution for various parameters in cross-platform complex transactions. Shoshi et al. (see \cite{shoshih}) proposed the volatility method and the duration method to capture the random behavior of the time series and analyzed the Bakken crude oil data through various machine learning and deep learning algorithms. Similar analysis for the yield data in commodity market is provided in \cite{ShHan}. Roberts et al. (see \cite{robertsm1, robertsm}) proposed a sequential hypothesis test that can detect the general jump size distribution. Their method was applicable to the crude oil price data set and improved the stochastic model using various machine and deep learning algorithms.

Fuzzy theory is a powerful tool to describe uncertain events (see \cite{Zadeh}), and has a wide range of applications in management science, decision science, and intelligent control. There are many randomness and ambiguities in the financial market, which are mainly reflected in the uncertainty of market changes, the incomplete symmetry of the information of the parties involved in a transaction, and the real-time fluctuation of commodity transaction data. Some scholars have tried to apply fuzzy theory to the field of financial research and have achieved some effective results. For example, Hatami-Marbini et al. (see \cite{hatamim}) used fuzzy numbers to describe the indicators and factors of stock performance, and obtained fuzzy distances to optimize stock portfolios. The combination of fuzzy theory and deep learning is used to predict changes in high-frequency financial data (see \cite{dengy}). Thavaneswaran et.al. (see \cite{Thavaneswaran}) implemented the fuzzy set theory to price binary options by using trapezoid, parabola, and adaptive fuzzy stock price maturity value. Zhang et. al. (see \cite{Zhanglh}) considered the clear probability mean of the fuzzy number, and obtained the clear probability mean option pricing formula in the jump diffusion model of fuzzy double index. The use of fuzzy theory has obvious advantages in financial data processing. Data processing involving fuzzy parameters can accurately describe the fuzzy changes of real market situation. 

For this paper, we apply fuzzy sets and machine learning to the analysis of the stock index. In Section \ref{sec2} we describe the BN-S model and its various modifications. We compare the different advantages of the classical and generalized BN-S models.  The financial data processing problem under fuzzy random uncertainty is introduced in Section \ref{sec3}. In Section \ref{sec4}, we analyze the S\&P 500 index data, with machine/deep learning and fuzzy theory. A brief conclusion is provided in Section \ref{sec5}.

\section{The classical BN-S model and its generalizations}
\label{sec2}

The Barndorff-Nielsen and Shephard (BN-S) model is a type of stochastic volatility model that is commonly used to describe the dynamic changes of asset prices. This is implemented to capture the response patterns of some stylized characteristics of financial asset time series in the financial market. The non-Gaussian Ornstein-Uhlenbeck (OU) process in the BN-S model is driven by an incremental L\'evy process, which is a random process of positive mean recovery. Consider a frictionless financial market, where a stock, and a risk-free asset with a fixed rate of return $r$, are traded on the horizon date $T$. The BN-S model assumes that the price process of the stock (or, commodity), $S=(S_t)_{t \ge 0}$, defined in a filtered probability space $(\Omega, \mathcal{F}, (\mathcal{F}_t)_{0 \le t \le T}, P)$, is given by:
\begin{equation}
  S_t=S_0\exp(X_t).
\label{1}
\end{equation}
The log-return $X_t$ is governed by:
\begin{equation}
  dX_t=(\mu+\beta \sigma^2_t)dt+\sigma_t d W_t + \rho d Z_{\lambda t},
\label{2}
     \end{equation}
where the parameters $\mu, \beta, \rho \in \mathbb{R}$, and $\rho\le 0$, and the variance process is given by:
\begin{equation}
 d\sigma^2_t=-\lambda \sigma^2_t + dZ_{\lambda t}, \quad \sigma^2_0 >0.
\label{3}
\end{equation}
In \eqref{3}, $\lambda \in \mathbb{R}$, and $\lambda > 0$. For the probability measure $P$, $W = (W_t)$ is the standard Brownian motion defined in the probability space. The process $Z = (Z_{\lambda t})$ is a subordinator. This is also known as the background driving L\'evy process (BDLP). It is assumed that the processes $W$ and $Z$ are independent, and $(\mathcal{F}_t)$ is a conventional augmentation of the filtering produced by $(W, Z)$.

There are several issues in the application of the classical BN-S model. Both logarithmic return and volatility (or, variance) contain a single BDLP, which makes them completely dependent on each other, leading to inaccurate volatility simulations. This absolute correlation also implies that the model will fail in a longer time frame, that may span for just a few days. The model cannot consistently capture the basic characteristics of the relevant time series. For historical data, the jump in volatility is not completely synchronized with the jump in stock prices. The volatility $\sigma_t$ usually cannot immediately respond to sudden fluctuations in a stock (or, commodity) price. This is one of the reasons why the classical BN-S model fail to work.

Some of these issues are addressed in the generalized BN-S model (see \cite{Sengupta1}). The new model simulates option prices and volatility in an interrelated but different way.
The generalized model assumes that $Z_t$ and $Z_t^*$ are two independent L\'evy subordinators with same (finite) variance. Then, there exists a L\'evy subordinate $d\widetilde{Z}_{\lambda t}$ independent of $W$, which is given by:
\begin{equation}
d{\bar Z_{\lambda t}} = \rho 'd{Z_{\lambda t}} + \sqrt {1 - {(\rho ') ^2}} dZ_{\lambda t}^*, \quad 0 \le \rho ' \le 1.
\label{(2.4)}
\end{equation}   
For the generalized model, the dynamics of $S_t$ is given by \eqref{1} and \eqref{2}, where $\sigma_t^2$ is given by:
\begin{equation}
d\sigma _t^2 =  - \lambda \sigma _t^2dt + d{\bar Z_{\lambda t}}, \quad \sigma _0^2 > 0. 
\label{(2.5)}
\end{equation}  
In \eqref{(2.5)}, the OU process $\bar Z =(\bar Z_{\lambda t})$ is correlated to the corresponding $Z$ in \eqref{3} and is also independent of $W$ subordination.

The classical BN-S model, even with the above generalization, fails to accommodate long-range dependence. In a recent paper, some major refinements are made to the BN-S model. This model is implemented in \cite{robertsm, sengupta4, shoshih}. For this refinement of the BN-S model, the log-return $X_t$, on some appropriate risk-neutral filtered probability space, is considered to be driven by a convex combination of two subordinators $Z$ and $Z^{(b)}$. This is written as:
\begin{equation}
 d{X_t} = (\mu  + \beta \sigma _t^2)dt + {\sigma _t}d{W_t} + \rho \left( {(1 - \theta )d{Z_{\lambda t}} + \theta dZ_{\lambda t}^{(b)}} \right),
\label{(2.6)}
\end{equation} 
where $\theta$ is a deterministic parameter, with $\theta \in [0,1]$, $\lambda >0$ is the proportional parameter at $t$, and $Z$ and $Z^{(b)}$ are independent L\'evy processes. We asssume that $Z^{(b)}$ corresponds to a subordinator with greater L\'evy intensity in comparison to the subordinator $Z$.  In addition, when the dynamics of $X_t$ is given by \eqref{(2.6)}, the variance process is given by:
\begin{equation}
d\sigma _t^2 =  - \lambda \sigma _t^2dt + (1 - \theta' )d{Z_{\lambda t}} + \theta'dZ_{\lambda t}^{(b)}, \quad \sigma _0^2 > 0.
\label{(2.7)}
\end{equation}
Similar to the expression \eqref{(2.5)}, $\theta' \in [0,1]$ is deterministic, and the processes $Z$ and  $Z^{(b)}$ are independent L\'evy processes. The sum of $(1-\theta)Z_{\lambda t}$ and $\theta Z_{\lambda t}^{(b)}$ is also a L\'evy process and is positively correlated with $Z$ and $Z^{(b)}$.

The integral variance is given by  $\sigma_I^2=\int_t^T \sigma_s^2ds$ in the time period $[t, T]$. Consequently, \eqref{(2.7)} provides: 
\begin{equation}
 \sigma_I^2= \epsilon (t,T)\sigma_t^2+\int_t^T \epsilon(s,T)((1-\theta')dZ_{\lambda t}+\theta' dZ_{\lambda t}^{(b)}), 
\label{(2.8)}
\end{equation}
where
\begin{equation*}
\epsilon (s,T)=(1-\exp(-\lambda(T-s)))/\lambda, \quad t\le s \le T.
\end{equation*}
Consequently, the continuous realized variance in the interval $[0, T]$ is given by:
\begin{equation}
 \sigma_R^2=  \frac{1}{T}\int_0^T \sigma_t^2dt+\rho^2(1-\theta)^2\lambda Var[Z_1]+\rho^2 \theta^2 \lambda Var[Z_1^{(b)}].
 \label{(2.9)} 
\end{equation}
We assume $\theta = \theta'$ for the convenience for the rest of the paper. If $J_Z$ and $J_Z^{(b)}$  are jump processes related to the subordinate $Z$ and  $Z^{(b)}$, respectively, and we define $J(s)=\int_0^s \int_{\mathbb{R}+}J_Z(\lambda d\tau ,dy)$, and $J^{(b)}_{(s)}=\int_0^s \int_{\mathbb{R}+}J_Z^{(b)}(\lambda d\tau ,dy)$, then for the log-return of the classical BN-S model:
\begin{equation}
Corr(X_t,X_s)= \frac{\int_0^s \sigma_\tau^2d\tau +\rho^2J(s)}{\sqrt{(\int_0^t \sigma_\tau^2 d \tau + t \rho^2 \lambda Var(Z_1))(\int_0^s \sigma_\tau^2 d \tau + s \rho^2 \lambda Var(Z_1))}}, \quad t>s.
\label{(2.10)}
\end{equation}
For the refined BN-S model the result becomes:
\begin{equation}
Corr(X_t,X_s)= \frac{\int_0^s \sigma_\tau^2d\tau +\rho^2(1-\theta)^2J(s)+\rho^2 \theta^2 J^{(b)}(s)}{\sqrt{\alpha(t)\alpha(s)}}, \quad t>s,
\label{(2.11)} 
\end{equation}
where $\alpha(\nu)=\int_0^\nu \sigma^2_\tau d\tau +\nu \rho^2 \lambda((1-\theta)^2Var(Z_1)+\theta^2Var(Z_1^{(b)})).$

In \eqref{(2.10)}, for a fixed $s$,  as $t$ increases $Corr (X_t, X_s)$ rapidly becomes small. It shows that the classical BN-S model is affected by time changes in the process of fitting random fluctuations, which leads to inaccurate volatility simulation. This rapid attenuation means that the model will be inaccurate in a longer time range. Consequently, the classical BN-S model is unable to accurately capture the basic characteristics of the relevant time series. However, due to the parameter $\theta$, and  the fact that $t$ always has an upper limit, for a fixed $s$, $Corr (X_t, X_s)$ in \eqref{(2.11)} will never become ``too small".  Hence, compared to the classical model, the generalized BN-S model is more efficient in extracting a deterministic component from a stochastic financial process. The refined model improves the long-term dependence problem of the classical model with a few parameter changes. At the same time, it provides dynamic characteristics with obvious advantages for analysis of financial time series data.

\section{Price representation based on fuzzy theory}
\label{sec3}

There are many uncertainties in the financial market. The advantage of treating the daily price of stock index as a fuzzy parameter is that it can accurately describe the range of daily price fluctuations. It can bypass the errors caused by some unreasonable data in the empirical data-set, and increase the accuracy and operability of the yield and risk quantification process. Fuzzy numbers are often used to describe uncertain information. The definition of a fuzzy set requires two components: a universe of discourse or domain, and a function. The function is called the membership function, which defines the ``degree" to which a particular element of the domain belongs to the set. Fuzzy sets are sets whose elements have degrees of membership. 

Let $S$ be the a universe of discourse or domain. We assume that $S$ is is a subset of $\mathbb{R}$.  We consider a fuzzy set $A \subset S$. There is a membership function $\mu_A: S \to  [0, 1]$ corresponding to each $x \in S$. The value of $\mu_A(x)$ represents degree of membership, that quantifies the grade of membership of the element in $S$ to the fuzzy set $A$. We form fuzzy parameters by associating $\mu_A(x)$ with real data to realize the quantification of the fuzzy environment in the financial market. For $l < m < r$, the general representation of the membership function $\mu$ can be written as:

$$
  \mu_A(x)= \left \{
       \begin{array}{ll}
  L(x),  \quad l \le x \le m,\\
 R(x),  \quad m \le x \le r.\\
     \end{array}
       \right.
 \eqno{(3.1)} $$
In the above expression, $L(x)$ is a right continuous increasing function, and $0 \le L(x) \le 1$; $R(x)$ is a left continuous decreasing function, and $0 \le R(x) \le 1$.
The value of membership (also know as ```confidence") $\alpha \in [0,1]$ is usually expressed as: $ A \alpha=\{x:\mu_A(x) > \alpha\} $, the $\alpha$ level of fuzzy set $A$ constitutes the set of all elements whose membership of $A$ is greater than or equal to $\alpha$ in the complete set.

\begin{figure}
\centering
{\includegraphics[width=4in,height=3in]{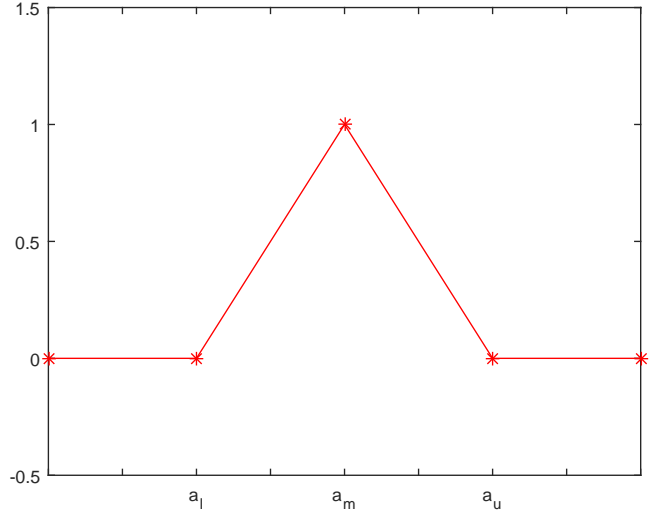}}
\caption{  \small{Triangular fuzzy number distribution}}
\end{figure}

As an example, triangular fuzzy number is one of the classical expressions of fuzzy number. This  is widely used in fuzzy evaluation system. The membership function $\mu_A(x)$ is used to show the degree to which the element $x$ belongs to the fuzzy set $A$. As shown in Figure 1, the triangular fuzzy number is a continuous convex function, composed of linear non-decreasing parts and non-increasing parts. Generally, the membership function of the triangular fuzzy number $A=(a_l, a_m, a_u)$, where $0 \le a_l \le a_m \le a_u \le 1$,  is expressed as follows:
$$
  \mu_A(x)= \left \{
       \begin{array}{ll}
  0, \quad \text{where} \quad  x \le a_l,\\
 \frac{ {x-a_l}}{ {a_m-a_l}}, \quad \text{where} \quad  a_l \le x \le a_m,\\
  \frac{ {a_u-x}}{ {a_u-a_m}}, \quad \text{where} \quad   a_m \le x \le a_u,\\
  0,  \quad \text{where} \quad    x \ge a_u.\\
     \end{array}
       \right.
 \eqno{(3.2)} $$
Here $a_l$ and $a_u$ are called the left-most and right-most values of fuzzy set $A$, respectively. They describe the lower and upper limits of the triangular fuzzy number $A$. Their difference indicates the fuzzy degree of fuzzy set $A$. The number $a_m$ is called the kernel of $A$ and it represents the most likely value of the triangular fuzzy number $A$. In particular, if $a_l=a_m=a_u$, then the fuzzy number degenerates into a real number. In addition (see \cite{Xu}), the $\alpha^-$ level set ($\alpha^-$ cut set) of $A$ is denoted as: $\tilde{A} =[A^L, A^R]$, where $A^L = (1-\alpha) a_l+\alpha a_m $, $A^R = (1-\alpha) a_u +\alpha a_m$. If both $A^L$ and $A^R$ are integrable, then the expectation $E(A)$ of set $A$ is given by: $E (A)=[(1-\lambda)a_l+a_m+\lambda a_u]/2$, where $0 \le \lambda \le 1$. The value of $\lambda$ depends on the importance of the influence of the fuzzy boundary.

In order to improve the volatility structure in the BN-S model, and to address the issue that some input parameters in the model are difficult to accurately estimate, we consider the daily closing price,  daily highest price, and  daily lowest price of a stock index. We use triangular fuzzy numbers to describe the fuzzy daily price. With the help of fuzzy theory, for the daily price variables of stock index, we can obtain the price $S=(s_l, s_m, s_u)$ in fuzzy form. The new fuzzy daily price is composed of three real numbers, which correspond to the lowest ($s_l$), closing ($s_m$) and highest ($s_u$) price of the empirical data on a particular date. From the empirical data of stock index price, we can get the triangular fuzzy number $S=(s_l, s_m, s_u)$, that represents corresponding fuzzy price. The three parameters in the fuzzy set correspond to the daily lowest, closing, and highest price, respectively. Consequently, we define the value of $\lambda$ to get fuzzy price expectation 
\begin{equation*}
E(S)=\frac{(1-\lambda)\text{Price}_{\min}+\text{Price}_{\text{close}}+ \lambda \text{Price}_{\max}}{2},
\end{equation*}
where $ S=(\text{Price}_{\min},\;\text{Price}_{\text{close}},\;\text{Price}_{\max})$. This will be used as an indicator of the daily fuzzy price for the next analysis.

\section{Numerical analysis based on machine/deep learning}
\label{sec4}

\subsection{Data description and fuzzy parameters}

In this section, we provide numerical examples to find the value of $\theta$ in the refined BN-S model \eqref{(2.6)}. We consider the S\&P 500 data set from November 1, 2010, to October 31, 2020, which contains $2518$ data-points (Data source: \url{https: / /finance. yahoo. com/}). Some statistics of the empirical data set over the given time-period are shown in Table 1. The daily closing price data, can be an indicator for the daily rise and fall of the S\&P 500 stock index. Figure 2 shows the annual plot of the close price, daily rise and fall, and volatility of the empirical data-set.

\begin{table}[h]
\caption{Properties of the empirical data set}
\begin{center}
\begin{tabular}{|c|c|c|c| p{4cm}|}
\hline
  & Daily Price Change & Daily Price Change \% & Daily Volatility Range \\ \hline
Mean & 0.83 &  0.04  & 22.36  \\
\hline
Median & 1.43 &  0.07  & 16.59  \\
\hline
Minimum & -228.62 &  -8.56  & 0\\
\hline
Maximum & 180.36 &  8.04  & 218.96\\
\hline
\end{tabular}
\end{center}
\end{table}

\begin{figure}[h]
\centering
{\includegraphics[width=6in,height=4in]{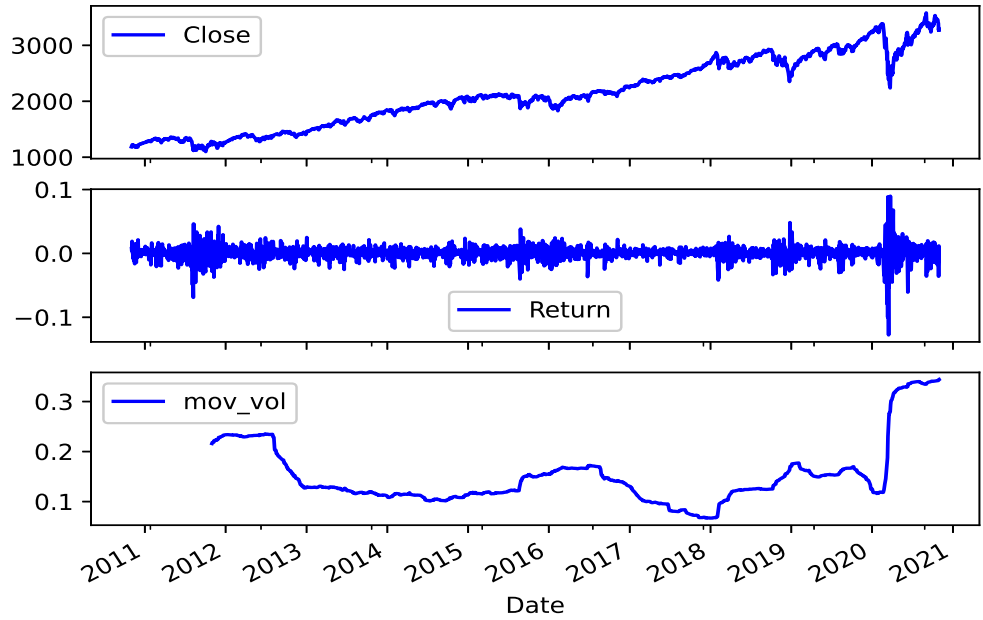}}
\caption{  \small{S\&P 500 close price, daily rise and fall and volatility}}
\end{figure}

Affected by many factors inside and outside the financial market, the price of stock index fluctuates many times within a day. In the process of looking for the deterministic component ($\theta$) in the stochastic time series, we hope to find a suitable daily price parameter to describe the fluctuation, which has both randomness and ambiguity. Data preprocessing using fuzzy parameters can describe the fuzzy situation of the range of price changes more accurately. We adjust the value of $\lambda$ in the triangular fuzzy number, so that the fuzzy price accurately describes different risk preferences, different market trends, and different investment objectives. For investors with different risk preferences, the value of $\lambda$ is different. Risk-averse investors are more likely to be affected by the lower limit of the fuzzy price boundary. Consequently, we assume that the degree of risk aversion of investors is inversely proportional to the value of $\lambda$. It is natural to assume that the $\lambda$ corresponding to a risk-neutral investor is $0.5$, and the aggressive risk pursuer corresponds to a higher value of $\lambda$. The value of $\lambda$ is also affected by the market environment. For example, in a bull market, the overall operating trend of a long market is upward, and the upper limit of the fuzzy boundary is more important. In that case, the value of $\lambda$ is typically larger than that in a bear market. Different values of $\lambda$ can also describe different investment objectives. 

For example, in the case of a put option, investors pay more attention to the falling price of the underlying asset (especially the lowest price). So $\lambda$ in the fuzzy price describing the volatility of the put option is generally between $0$ and $0.5$. With the price drop, the value of $\lambda$ tends to $0$. On the other hand, a call option is just the opposite. It is important in investment decisions to estimate the price increase (especially the highest price) of the underlying asset. So $\lambda$ in the fuzzy price describing the volatility of the call option is generally between $0.5$ and $1$. With the price growth, the value of $\lambda$ tends to $1$.

Generally, the daily closing price is the volume-weighted average price of all transactions one minute before the last transaction on that day. The daily highest price and daily lowest price describe the degree of price change. In this paper, we use triangular fuzzy numbers to calculate the daily fuzzy price of the stock index option based on three variables: daily lowest price, daily highest price, and daily closing price. In Table 2, we list the feature estimators of fuzzy price data when $\lambda$ takes different values. Figure 3 provides a time series chart of the fuzzy price of S\&P 500 stock index. We take $\lambda=0.5$ for machine learning and deep learning on the empirical data in the following section.

\begin{table}[h!]
\caption{Fuzzy price of the empirical data set}
\begin{center}
\begin{tabular}{|c|c|c|c|c| p|}
\hline
  & Daily fuzzy & Daily fuzzy Price 	& Daily fuzzy Price & Daily fuzzy Price \\
& Price parameters    &  $(\lambda=0.3)$	& $(\lambda=0.5)$	&  $(\lambda=0.7)$\\
\hline
Mean & (2118.59, 2130.65, 2140.96) &	2127.98 &	2130.21 &	2132.45  \\
\hline
Median & (2066.58, 2078.18,2085.19)	& 2075.17 &	2077.03  &	2078.89   \\
\hline
Minimum & (1074.77, 1099.23, 1125.12) &	1094.55  &	1099.59 &	1104.62 \\
\hline
Maximum & (3535.23, 3580.84, 3588.11) &	3565.97  &	3571.26 &	3576.54 \\
\hline
\end{tabular}
\end{center}
\end{table}

\begin{figure}[h]
\centering
{\includegraphics[width=6in,height=3in]{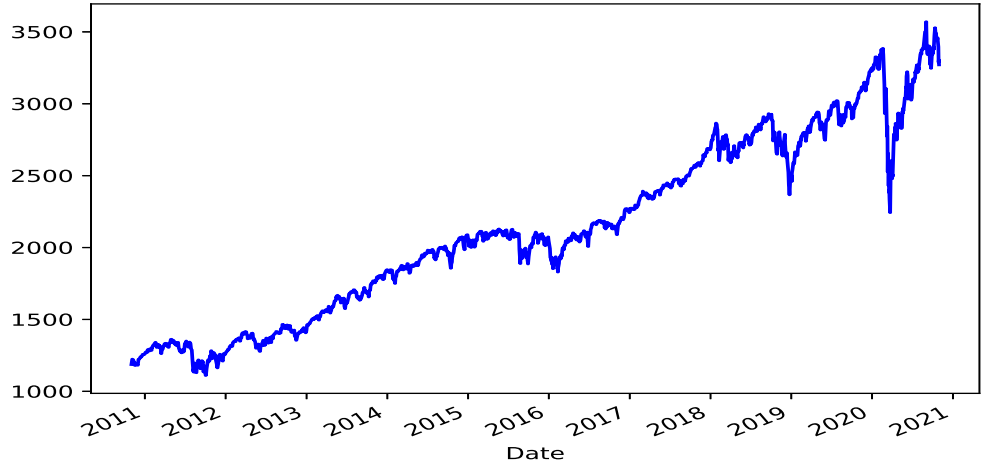}}
\caption{  \small{S\&P 500 fuzzy price (November 1, 2010 to October 30, 2020)}}
\end{figure}

\subsection{Machine learning and deep learning}
We index the available fuzzy price data by date. Based on the attributes of the data set, we construct a machine learning classification problem and provide quantitative decision support for estimating the value of $\theta$ with reasonable accuracy. Specific steps are as follows:
\begin{itemize}
\item[Step 1]: From the available data, we arrange the preprocessed fuzzy price data in the order of date (from November 1, 2010 to October 31, 2020).
\item[Step 2]: Explore the trend and distribution of fuzzy price changes over time, and visualize the data structure by month and year. Figure 4 provides moving averages of fuzzy prices in various time periods ($5$ days, $42$ days, and $252$ days) to show the trend of S\&P 500 index price. Figures 5 and 6 show the dispersion and distribution of annualized data through box plots and histograms, respectively. In Figure 7, we show the bar graph of monthly data.
\item[Step 3]: Based on Step 2, we calculate the daily changes of fuzzy prices and list these changes. Figures 8 and 9 are histograms of the daily change and daily change percentage of fuzzy price, respectively.
\item[Step 4]: We continue to quantify price volatility. Based on Step 2, we calculate the realized volatility and the realized volatility return of the fuzzy price sequence. Figure 10 and Figure 13 provide the heat map and line graph of the realized volatility, respectively. Figures 11 and 12 show the heat map and line diagram of the realized volatility return, respectively.
\item[Step 5]: We define the threshold value of the fuzzy price change percentage as $C$. We look for the date(s) when the fluctuation is lower than the previous day's $C$-value. We denote those dates as the dates with ``big jump" of the fuzzy price. (For example, if $C=1$, the date when the fuzzy price is 1\% lower than the previous business day is a ``big jump" day).
\item[Step 6]: Referring to the figures and tables in the above steps, we summarize the fluctuation characteristics of the data set and divide the empirical data. This step is aimed at creating a new data structure from the existing data set. We take the fuzzy price change percentage for $10$ consecutive days as an array with $10$ elements in one row. After that, we augment the next row with the corresponding values for the next $10$-days, and so on. The matrix will be:
$$
\begin{pmatrix}
a_1, a_2, a_3, \cdots \cdots , a_{10} \\
a_2, a_3, a_4, \cdots \cdots , a_{11}  \\
a_3, a_4, a_5, \cdots \cdots , a_{12} \\
\cdots \cdots \\
a_{2509}, a_{2510}, a_{2511}, \cdots \cdots , a_{2518}
\end{pmatrix}.
$$
\item[Step 7]:  We add a new target column $\theta$ to the new data frame. If there are at least two ``big jumps" in the next $10$ days, we set $\theta=1$. Otherwise, we set $\theta=0$.
\item[Step 8]: We run various machine learning and deep learning algorithms to classify the new matrix data. The input is the daily change percentage of the fuzzy price for $10$ consecutive days, and the output is the $\theta$ value ($0$ or $1$) in the target column.
\end{itemize}

\begin{figure}
\centering
{\includegraphics[width=6in,height=2.5in]{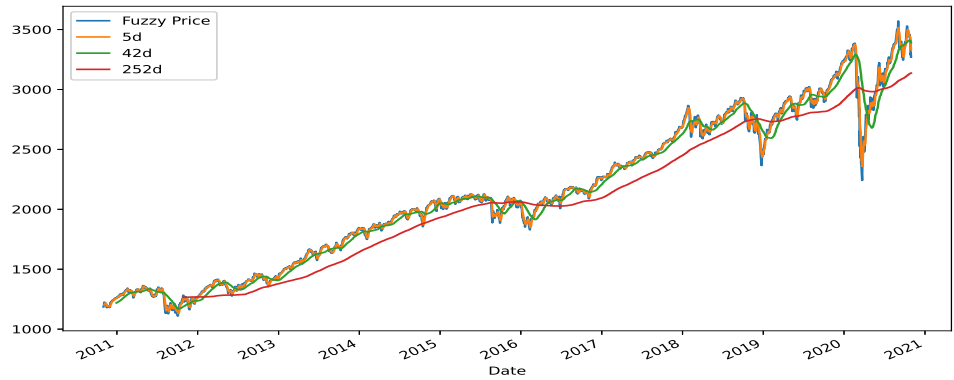}}
\caption{  \small{Moving average for the fuzzy price}}
\end{figure}
\begin{figure}
\centering
{\includegraphics[width=6in,height=2.5in]{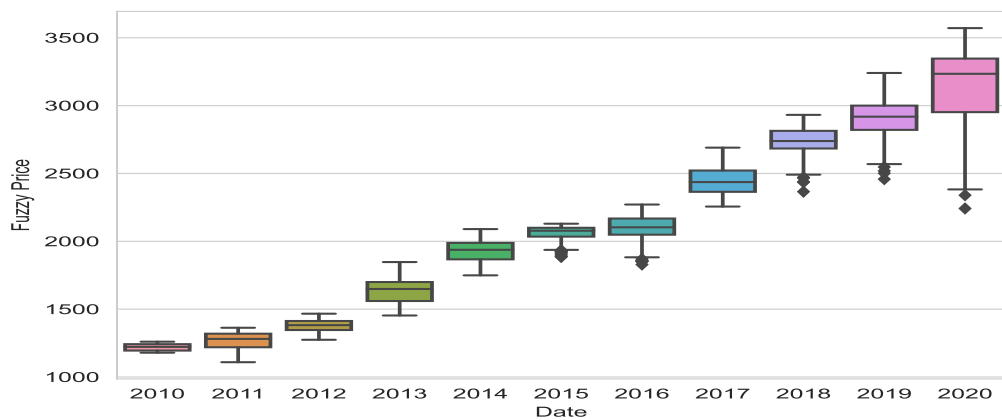}}
\caption{  \small{Yearly boxplot for the fuzzy price}}
\end{figure}
\begin{figure}
\centering
{\includegraphics[width=6in,height=2.5in]{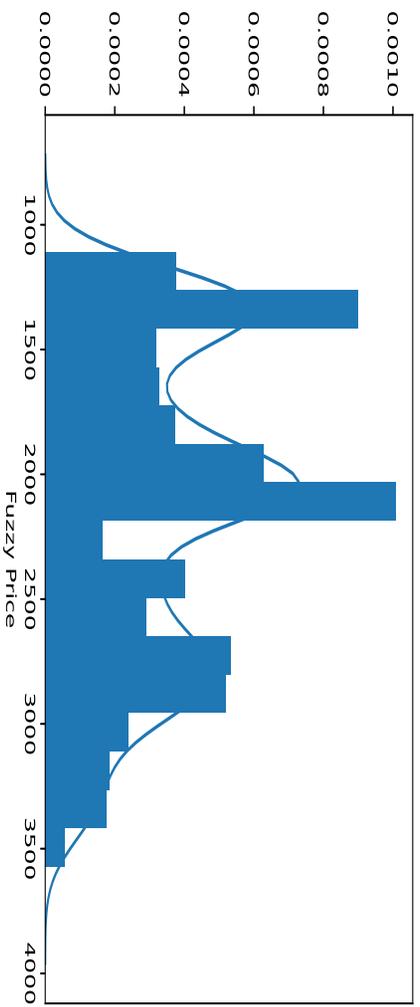}}
\caption{  \small{ Distribution plot for fuzzy price}}
\end{figure}
\begin{figure}
\centering
{\includegraphics[width=6in,height=2.5in]{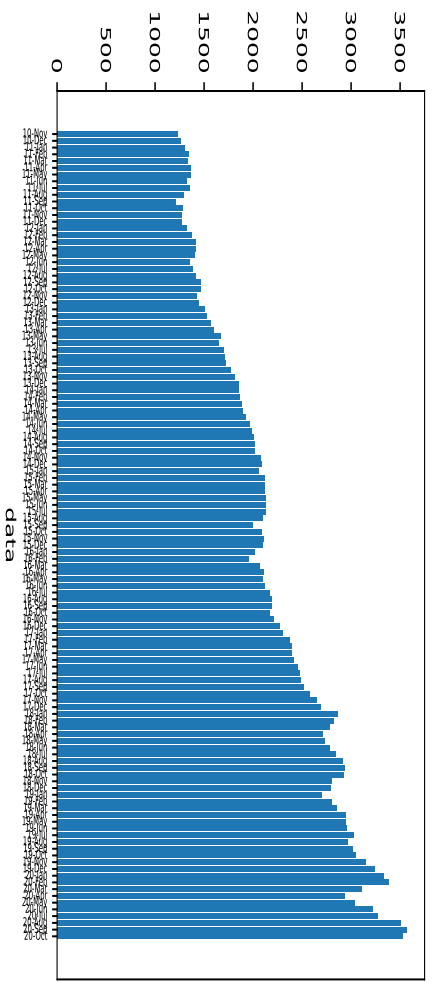}}
\caption{  \small{ Bar chart for fuzzy price}}
\end{figure}

\begin{figure}
\centering
{\includegraphics[width=6in,height=3in]{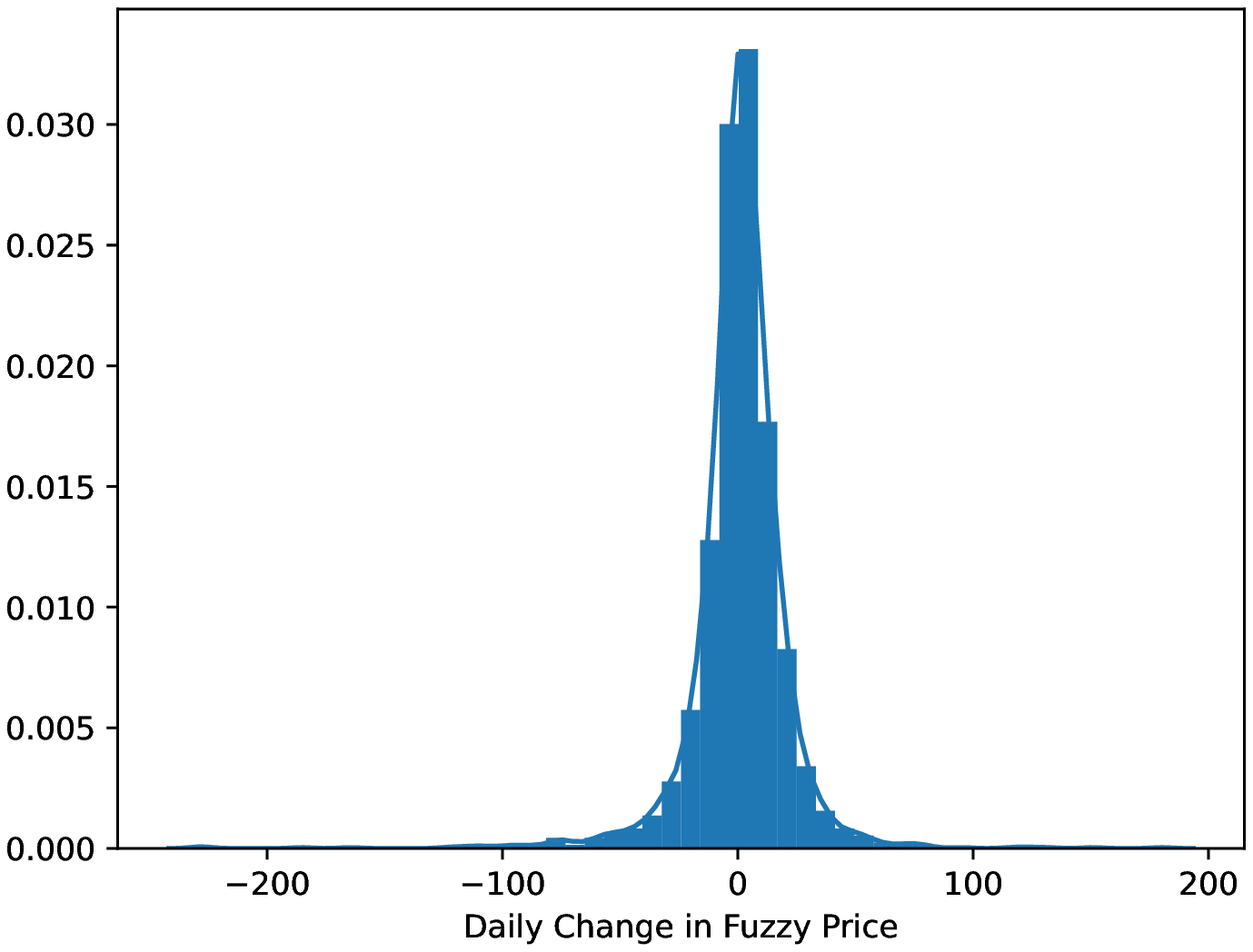}}
\caption{  \small{Histogram for daily change in fuzzy price}}
\end{figure}

\begin{figure}
\centering
{\includegraphics[width=6in,height=3in]{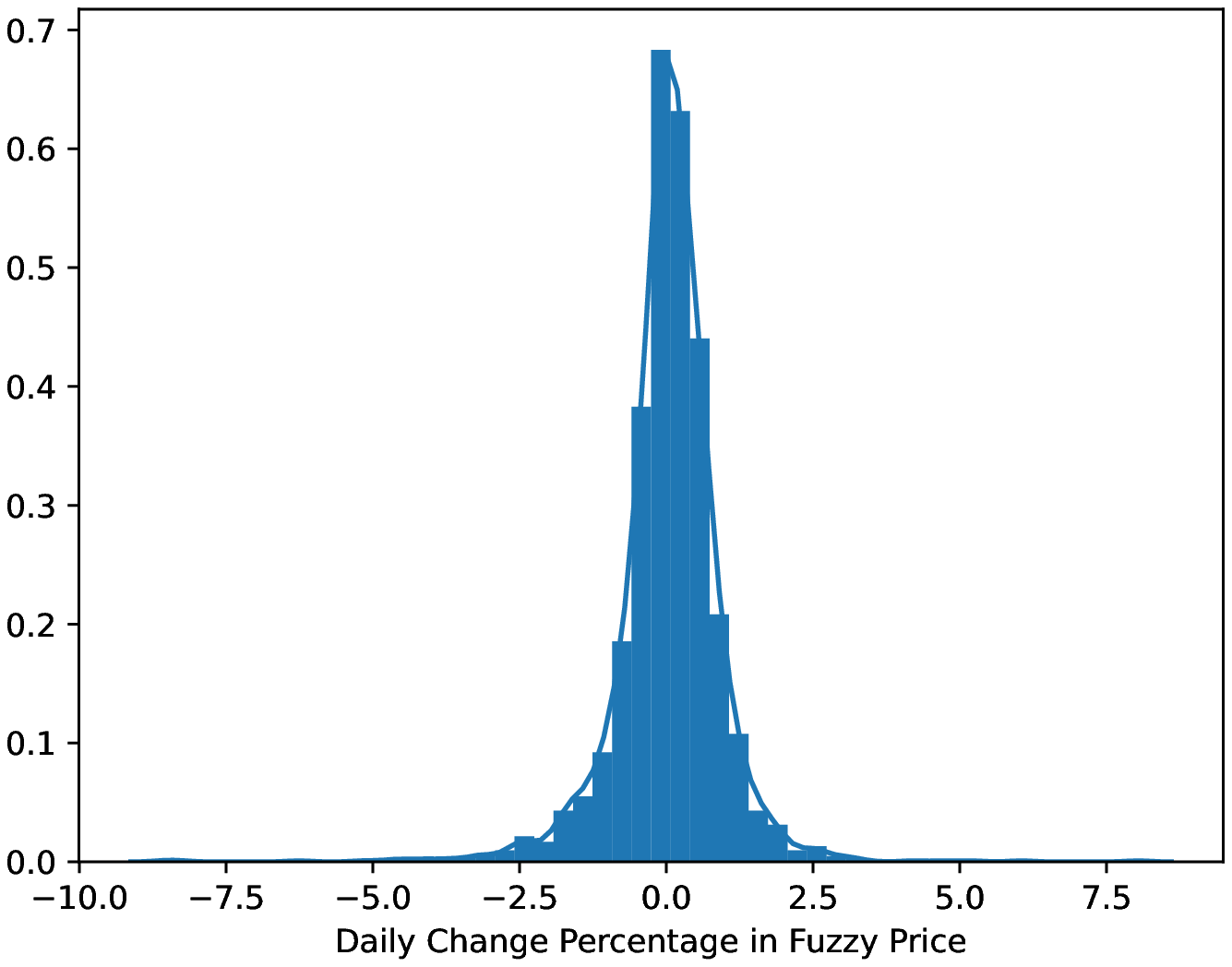}}
\caption{  \small{ Histogram for daily change percentage in fuzzy price}}
\end{figure}

\begin{figure}
\centering
{\includegraphics[width=4in,height=3in]{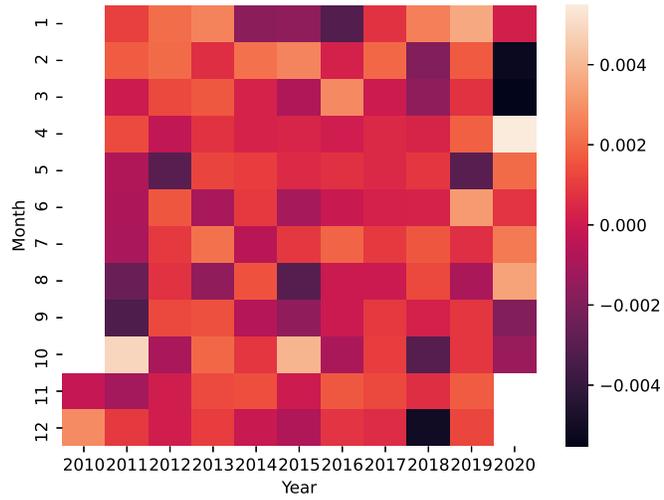}}
\caption{  \small{ Heatmap for the realized volatility of the fuzzy price over ten years}}
\end{figure}

\begin{figure}
\centering
{\includegraphics[width=4in,height=3in]{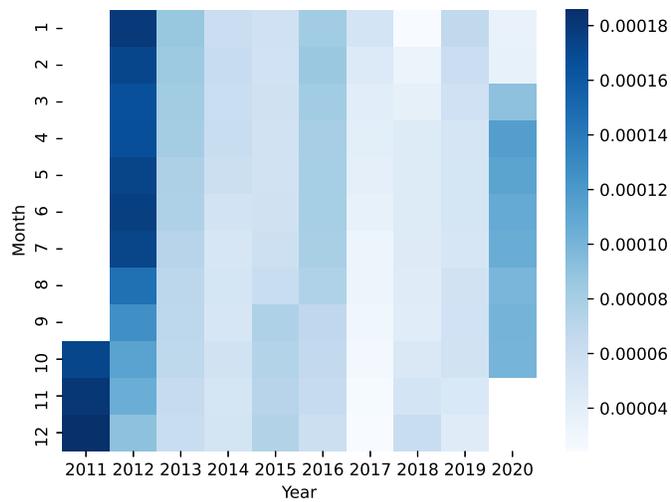}}
\caption{  \small{ Heatmap for the realized volatility return in percentage over the ten years for the fuzzy price}}
\end{figure}

\begin{figure}
\centering
{\includegraphics[width=6in,height=3in]{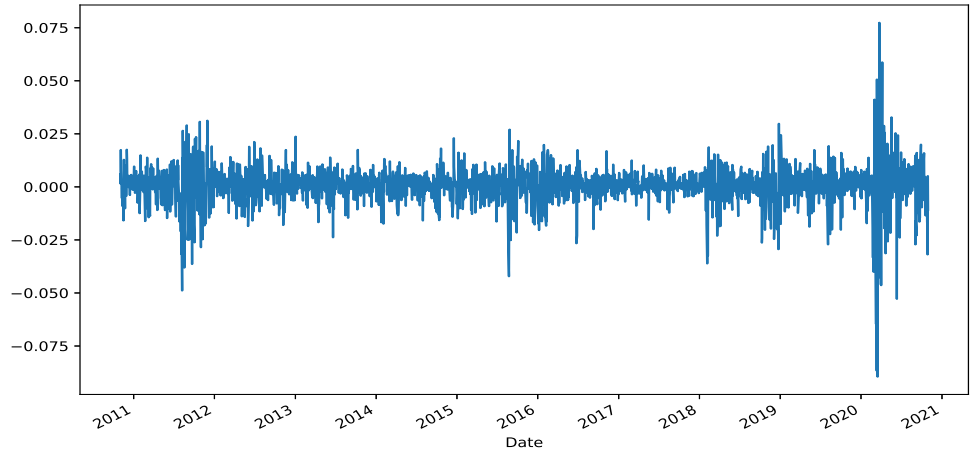}}
\caption{  \small{ Line Plot for the realized volatility of the fuzzy price}}
\end{figure}

\begin{figure}
\centering
{\includegraphics[width=7in,height=3in]{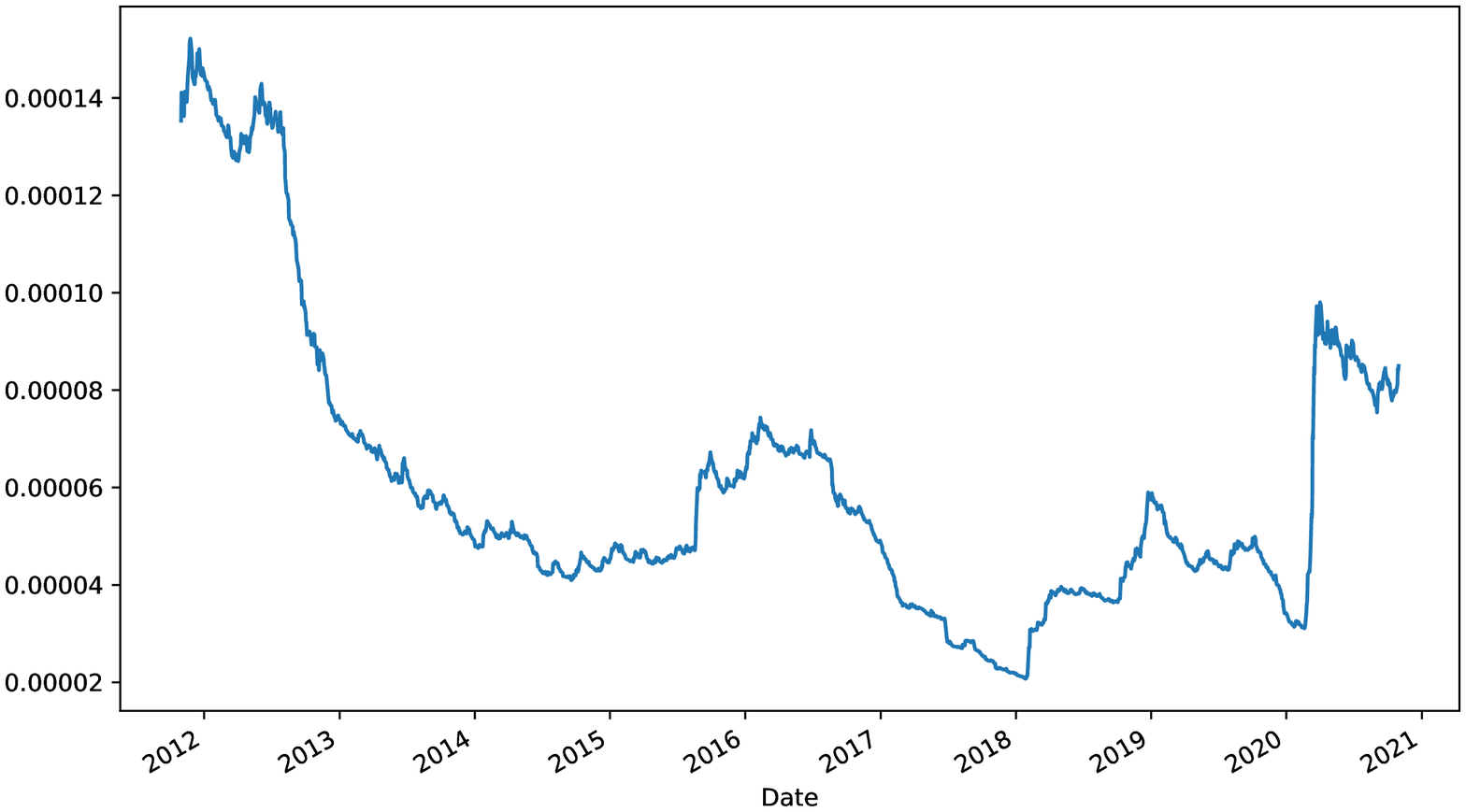}}
\caption{  \small{ Line plot for the realized volatility return in percentage for the fuzzy price}}
\end{figure}

It is worth noting that we can improve the result by adjusting the value of $C$ in Step 5. In addition, adjusting the data division in Step 6, we can also improve the validity of the results. The various machine learning and deep learning models involved in Step 8 provide $\theta$ values between $0$ and $1$. By implementing the above steps, we can find $\theta$ with reasonable accuracy and apply it to the refined BN-S model. The specific algorithm and calculation results in Step 8 are  are summarized below: 

\begin{itemize}
\item[(A)] Logistic regression:  Logistic regression is a well-known method in machine learning. It is often used for two-result classification. 
\item[(B)] Decision tree:  As a supervised classification algorithm, decision tree is a process of generating decision results (or tree diagram) based on historical experience (or training set). We perform attribute selection on fuzzy price data, measure and determine the topological structure between each characteristic attribute, construct a decision tree, and then find the value of $\theta$ through the overall optimal decision result.
\item[(C)] Random Forest: Random forest is a classifier in machine learning that contains many decision trees and can be effectively run on large data sets. Its output is determined by the mode of the category output by the individual tree. We divide the fuzzy price data set into a training set and a testing  set, instantiate the model, and use the standardized data to fit the model. After that, we measure the accuracy of the model through the training data, and then estimate the $\theta$ value.
\item[(D)] Neural network: Neural network is the basic algorithm of deep learning. It performs distributed and parallel information processing by constructing multiple ``neurons" to form a multilayer network. We build a neural network with two hidden layers and an output layer to estimate $\theta$. 
\item[(E)] Long and short-term memory neural network (LSTM): LSTM is a special recurrent neural network (RNN) that has the characteristics of maintaining long-term memory of information. We put the implementation of LSTM in the LstmLayer class, use the forward method to achieve forward calculation, and use the backward method to achieve back propagation. The activation function of the gate is the sigmoid function, and the output activation function is tanh.
\item[(F)] Batch normalizer (BN) in LSTM network: Batch normalizer solves the problem of vanishing-gradient or gradient-explosion by adjusting the input of the activation function. It also helps to improve the training speed of the LSTM network.
\end{itemize}

In the result tables (Tables 3 to 12), we provide classification reports based on the above algorithms. Some  of the times, the results of machine learning are not completely accurate. In order to observe the feasibility of each algorithm, we use ``precision" to represent the accuracy of all (i.e., $\theta = 1$ and $\theta = 0$) estimation results. This can be specifically quantified as the ratio of the number of accurately predicted $\theta = 1$ (or $\theta = 0$) to the number of all $\theta = 1$ (or, $\theta = 0$) prediction results. ``Recall" is used to describe the efficiency of $\theta = 1$ (or, $\theta = 0$) being accurately predicted. It can be specifically quantified as the ratio of the number of accurately predicted $\theta = 1$ (or, $\theta = 0$) to the actual number of $\theta = 1$ (or, $\theta = 0$). The values of ``precision" and ``recall" are positively correlated with the accuracy of the algorithm's prediction. We compute ``$f1$-score" (the harmonic mean of ``precision" and ``recall") to evaluate the prediction effect of each algorithm. In the following tables, ``support" means the number of samples in the calculation.

\begin{table}[h]
\center
\caption{Evaluation of calculation results within 1 year\protect\\ (Train set: 11/01/2019-05/13/2020, Test set: 05/14/2020-10/30/2020)}
 \setlength{\tabcolsep}{5mm}
 \begin{tabular}{rcccccc}
  \toprule
   & (A) & (B) &(C) &(D) &(E) &(F)  \\
  \midrule
precision $\theta =0$   & 0.73 & 0.75 &0.79 &0.62 &0.71 &0.51  \\
recall $\theta =0$      & 0.80 & 0.48 &0.64 &0.28 &0.43 &0.7  \\
f1-score $\theta =0$    & 0.76 & 0.59 &0.71 &0.39 &0.53 &0.35  \\
support $\theta =0$     & 75 & 75 &75 &75 &75 &75  \\
precison $\theta =1$    & 0.44 & 0.36 &0.44 &0.28 &0.33 & 0.21  \\
recall $\theta =1$      & 0.35 & 0.65 &0.62 & 0.62 & 0.62 &  0.44  \\
f1-score $\theta =1$    & 0.39 & 0.46 & 0.51 & 0.39 & 0.43 &0.29  \\
support $\theta =1$     & 34 & 34 &34 &34 &34& 34  \\
  \bottomrule
 \end{tabular}
\end{table}

\begin{table}
\center
\caption{Evaluation of calculation results within 2 year \protect\\ (Train set: 11/01/2018-05/13/2020, Test set: 05/14/2020-10/30/2020)}
 \setlength{\tabcolsep}{5mm}
 \begin{tabular}{rcccccc}
  \toprule
& (A) & (B) &(C) &(D) &(E) &(F)  \\
  \midrule
precision $\theta =0$   & 0.71 &0.67  &0.76 &0.76&0.66 &0.81 \\
recall $\theta =0$      & 0.89 & 0.49 &0.77 &0.68 &0.36 &0.59  \\
f1-score $\theta =0$    & 0.79 & 0.57 &0.77 &0.72 &0.47 &0.68  \\
support $\theta =0$     & 75 & 75 &75 &75 &75 &75  \\
precison $\theta =1$    & 0.47 & 0.30 &0.48 &0.43 &0.29 &0.44  \\
recall $\theta =1$      & 0.21 & 0.47 &0.47 &0.53 &0.59 &0.71  \\
f1-score $\theta =1$    & 0.29 & 0.36 &0.48 &0.47 &0.39 &0.54  \\
support $\theta =1$     & 34 & 34 &34 &34 &34& 34  \\
  \bottomrule
 \end{tabular}
\end{table}

\begin{table}
\center
\caption{Evaluation of calculation results within 3 year \protect\\ (Train set: 11/01/2017-07/29/2019, Test set: 07/30/2019-10/30/2020)}
 \setlength{\tabcolsep}{5mm}
 \begin{tabular}{rcccccc}
  \toprule
 & (A) & (B) &(C) &(D) &(E) &(F)  \\
  \midrule
precision $\theta =0$   & 0.71& 0.73 &0.76 &0.71 &0.76 &0.75  \\
recall $\theta =0$      & 0.93 & 0.64 &0.76 &0.70 &0.69&0.57  \\
f1-score $\theta =0$    & 0.80 & 0.68 &0.76 &0.71 &0.72 &0.65  \\
support $\theta =0$     & 203 & 203 &203 &203 &203 &203  \\
precison $\theta =1$    & 0.66 & 0.44 &0.54 &0.44&0.50 &0.44  \\
recall $\theta =1$      & 0.27 & 0.54 &0.55 &0.45 &0.58 &0.63   \\
f1-score $\theta =1$    & 0.39 & 0.48 &0.54 &0.45 &0.54 &0.52  \\
support $\theta =1$     & 106 & 106 &106 &106 &106 &106  \\
  \bottomrule
 \end{tabular}
\end{table}

\begin{table}
\center
\caption{Evaluation of calculation results within 4 year \protect\\ (Train set: 11/01/2016-07/29/2019, Test set: 07/30/2019-10/30/2020)}
 \setlength{\tabcolsep}{5mm}
 \begin{tabular}{rcccccc}
  \toprule
& (A) & (B) &(C) &(D) &(E) &(F)  \\
  \midrule
precision $\theta =0$   & 0.71& 0.70 &0.77 &0.68 &0.77 &0.84  \\
recall $\theta =0$      & 0.93 & 0.68 &0.79 &0.78 &0.61&0.72  \\
f1-score $\theta =0$    & 0.80 & 0.70 &0.78 &0.73 &0.68 &0.78  \\
support $\theta =0$     & 203 & 203 &203 &203 &203 &203  \\
precison $\theta =1$    & 0.66 & 0.45 &0.58 &0.41&0.47 &0.58  \\
recall $\theta =1$      & 0.25 & 0.50 &0.54 &0.29 &0.65 &0.74   \\
f1-score $\theta =1$    & 0.37 & 0.47 &0.56 &0.34 &0.54 &0.65  \\
support $\theta =1$     & 106 & 106 &106 &106 &106 &106  \\
  \bottomrule
 \end{tabular}
\end{table}

\begin{table}
\center
\caption{Evaluation of calculation results within 5 year \protect\\ (Train set: 11/01/2015-10/09/2018, Test set: 10/10/2018-10/30/2020)}
 \setlength{\tabcolsep}{5mm}
 \begin{tabular}{rcccccc}
  \toprule
& (A) & (B) &(C) &(D) &(E) &(F)  \\
  \midrule
precision $\theta =0$   & 0.70& 0.73 &0.76 &0.70 &0.76 &0.71  \\
recall $\theta =0$      & 0.97 & 0.63 &0.76 &0.87 &0.67&0.64  \\
f1-score $\theta =0$    & 0.81 & 0.68 &0.76 &0.77 &0.71 &0.67  \\
support $\theta =0$     & 336 & 336 &336 &336 &336 &336  \\
precison $\theta =1$    & 0.74 & 0.43 &0.54 &0.51&0.48 &0.42  \\
recall $\theta =1$      & 0.18 & 0.54 &0.54 &0.27 &0.59 &0.50   \\
f1-score $\theta =1$    & 0.29 & 0.48 &0.54 &0.35 &0.53 &0.45  \\
support $\theta =1$     & 173 & 173 &173 &173 &173 &173  \\
  \bottomrule
 \end{tabular}
\end{table}

\begin{table}
\center
\caption{Evaluation of calculation results within 6 year \protect\\ (Train set: 11/01/2014-10/09/2018, Test set: 10/10/2018-10/30/2020)}
 \setlength{\tabcolsep}{5mm}
 \begin{tabular}{rcccccc}
  \toprule
 & (A) & (B) &(C) &(D) &(E) &(F)  \\
  \midrule
precision $\theta =0$   & 0.69& 0.70 &0.75 &0.71 &0.78 &0.75  \\
recall $\theta =0$      & 0.99 & 0.64 &0.70 &0.83 &0.71&0.56  \\
f1-score $\theta =0$    & 0.82 & 0.67 &0.72 &0.77 &0.75 &0.64  \\
support $\theta =0$     & 336 & 336 &336 &336 &336 &336  \\
precison $\theta =1$    & 0.90 & 0.41 &0.48 &0.51&0.53 &0.43  \\
recall $\theta =1$      & 0.15 & 0.47 &0.55 &0.34 &0.62 &0.64   \\
f1-score $\theta =1$    & 0.26 & 0.44 &0.51 &0.41 &0.57 &0.51  \\
support $\theta =1$     & 173 & 173 &173 &173 &173 &173  \\
  \bottomrule
 \end{tabular}
\end{table}

\begin{table}
\center
\caption{Evaluation of calculation results within 7 year \protect\\ (Train set: 11/01/2013-12/21/2017, Test set: 12/22/2017-10/30/2020)}
 \setlength{\tabcolsep}{5mm}
 \begin{tabular}{rcccccc}
  \toprule
 & (A) & (B) &(C) &(D) &(E) &(F)  \\
  \midrule
precision $\theta =0$   & 0.71& 0.74 &0.75 &0.69 &0.70 &0.71 \\
recall $\theta =0$      & 0.98 & 0.69 &0.88 &0.88&0.66&0.66  \\
f1-score $\theta =0$    & 0.82 & 0.72 &0.81&0.77 &0.68 &0.68  \\
support $\theta =0$     & 481 & 481 &481 &481 &481 &481  \\
precison $\theta =1$    & 0.77 & 0.43 &0.59 &0.38&0.35 &0.38  \\
recall $\theta =1$      & 0.16 & 0.49 &0.38 &0.16 &0.39 &0.43   \\
f1-score $\theta =1$    & 0.26 & 0.46 &0.46 &0.23 &0.37 &0.40  \\
support $\theta =1$     & 228 & 228 &228 &228 &228 &228  \\
  \bottomrule
 \end{tabular}
\end{table}

\begin{table}
\center
\caption{Evaluation of calculation results within 8 year \protect\\ (Train set: 11/01/2012-12/21/2017, Test set: 12/22/2017-10/30/2020)}
 \setlength{\tabcolsep}{5mm}
 \begin{tabular}{rcccccc}
  \toprule
 & (A) & (B) &(C) &(D) &(E) &(F)  \\
  \midrule
precision $\theta =0$   & 0.71& 0.74 &0.75 &0.70 &0.75 &0.74 \\
recall $\theta =0$      & 0.98 & 0.68 &0.84 &0.87&0.64&0.73  \\
f1-score $\theta =0$    & 0.82 & 0.70 &0.79&0.77 &0.69 &0.74  \\
support $\theta =0$     & 481 & 481 &481 &481 &481 &481  \\
precison $\theta =1$    & 0.75 & 0.42 &0.54 &0.43&0.42 &0.45  \\
recall $\theta =1$      & 0.14 & 0.49 &0.40 &0.21 &0.56 &0.46   \\
f1-score $\theta =1$    & 0.24 & 0.45 &0.46 &0.28 &0.48 &0.45  \\
support $\theta =1$     & 228 & 228 &228 &228 &228 &228  \\
  \bottomrule
 \end{tabular}
\end{table}

\begin{table}
\center
\caption{Evaluation of calculation results within 9 year \protect\\ (Train set: 11/01/2011-10/13/2016, Test set: 10/14/2016-10/30/2020)}
 \setlength{\tabcolsep}{5mm}
 \begin{tabular}{rcccccc}
  \toprule
 & (A) & (B) &(C) &(D) &(E) &(F)  \\
  \midrule
precision $\theta =0$   & 0.79& 0.84 &0.83 &0.78 &0.79 &0.79 \\
recall $\theta =0$      & 0.99 & 0.72 &0.89 &0.88&0.75&0.76  \\
f1-score $\theta =0$    & 0.88 & 0.77 &0.86&0.83 &0.77 &0.77  \\
support $\theta =0$     & 776 & 776 &776 &776 &776 &776  \\
precison $\theta =1$    & 0.79 & 0.36 &0.52 &0.31&0.29 &0.29  \\
recall $\theta =1$      & 0.13 & 0.54 &0.40 &0.18 &0.33 &0.33   \\
f1-score $\theta =1$    & 0.22 & 0.43 &0.46 &0.23 &0.31 &0.31  \\
support $\theta =1$     & 233 & 233 &233 &233 &233 &233  \\
  \bottomrule
 \end{tabular}
\end{table}

\begin{table}
\center
\caption{Evaluation of calculation results within 10 year \protect\\ (Train set: 11/01/2010-10/13/2016, Test set: 10/14/2016-10/30/2020)}
 \setlength{\tabcolsep}{5mm}
 \begin{tabular}{rcccccc}
  \toprule
& (A) & (B) &(C) &(D) &(E) &(F)  \\
  \midrule
precision $\theta =0$   & 0.80& 0.80 &0.83 &0.79 &0.83 &0.85 \\
recall $\theta =0$      & 0.98 & 0.71 &0.86 &0.78&0.64&0.61  \\
f1-score $\theta =0$    & 0.88 & 0.75 &0.85&0.79 &0.72 &0.71  \\
support $\theta =0$     & 776 & 776 &776 &776 &776 &776  \\
precison $\theta =1$    & 0.72 & 0.29 &0.48 &0.31&0.32 &0.33  \\
recall $\theta =1$      & 0.18 & 0.39 &0.43 &0.32 &0.56 &0.65   \\
f1-score $\theta =1$    & 0.29 & 0.33 &0.45 &0.31 &0.41 &0.44  \\
support $\theta =1$     & 233 & 233 &233 &233 &233 &233  \\
  \bottomrule
 \end{tabular}
\end{table}

In these tables  (Tables 3 to 12), we show the dynamic estimation results of the value of $\theta$. Different algorithms have different prediction effects on the same data set. But different combinations of results and accuracy provide more decision support indicators for us to determine the value of $\theta$. According to the fuzzy data of stock index during different time spans, machine learning and deep learning algorithms can update the predicted value of $\theta$ in real time, and increase the accuracy of the prediction. It helps us find the deterministic component in the stochastic dynamics of stock index. Once $\theta$ is determined, we can apply it to the refined BN-S model introduced in Section \ref{sec2}.

\section{Conclusions}
\label{sec5}
The refined BN-S model inherits many advantages of the classical BN-S model in analyzing financial price fluctuations. In addition, it efficiently addresses some issues with the classical BN-S model. For example the refined model incorporates long-term dependence. This paper considers an application of the refined BN-S model in the context of stock index. Based on machine learning algorithms and fuzzy theory, deterministic component $\theta$ is extracted from completely random time series fluctuations of a financial stock index. Data preprocessing involving fuzzy parameters can more accurately describe the range of fuzzy changes in market prices. Fuzzy prices containing different risk preferences, different market trends, and different investment objectives help us find a more suitable $\theta$. Machine and deep learning algorithms are shown to help extract the effective information hidden in financial data. We apply various data-science-based techniques to identify $\theta$. Through the analysis of fuzzy price data, the fitting of random fluctuations is realized. The analysis is shown to be helpful to determine appropriate value of $\theta$.

The development and application of stochastic model introduced in this paper can optimize the function of traditional model, realize accurate dynamic fluctuation analysis, and enrich the theoretical basis of financial risk management. Future research will continue to focus on the application of the refined BN-S model in financial market analysis.


\section*{Acknowledgments}
This work is supported in part by the National Key Research and Development Program of China (2017YFB1401801), National Natural Science Foundation of China (71774042, 71532004) and China Scholarship Council (201906120273).

\end{document}